\documentstyle[12pt,epsf]{article}
\begin{document}
\catcode`@=11
\def\marginnote#1{}
%
\newcommand{\newc}{\newcommand}

\newc{\be}{\begin{equation}}
\newc{\ee}{\end{equation}}
\newc{\bea}{\begin{eqnarray}}
\newc{\eea}{\end{eqnarray}}

\newc{\gsim}{\lower.7ex\hbox{$\;\stackrel{\textstyle>}{\sim}\;$}}
\newc{\lsim}{\lower.7ex\hbox{$\;\stackrel{\textstyle<}{\sim}\;$}}
\newc{\thw}{\theta_W}
\newc{\ra}{\rightarrow}
\newc{\VEV}[1]{\langle #1 \rangle}
\newc{\hc}{{\it h.c.}}
\newc{\ie}{{\it i.e.}}
\newc{\etal}{{\it et al.}}
\newc{\eg}{{\it e.g.}}
\newc{\etc}{{\it etc.}}

\newc{\msbar}{\overline{\rm MS}}

\newc{\vrot}{v_{\rm rot}(r)}
\newc{\rhocrit}{\rho_{crit}}
\newc{\rhochi}{\rho_{\chi}}
\newc{\ev}{~{\rm eV}}     \newc{\kev}{~{\rm keV}}
\newc{\mev}{~{\rm MeV}}   \newc{\gev}{~{\rm GeV}}
\newc{\tev}{~{\rm TeV}}
\newc{\abund}{\Omega h^2_0}
\newc{\mx}{M_{GUT}}
\newc{\gx}{g_{GUT}}	\newc{\alphax}{\alpha_{\rm GUT}}
\newc{\msusy}{M_{SUSY}}
\newc{\msusyeff}{M^{eff}_{\rm SUSY}}
\newc{\ewmgg}{SU(2)_L\times U(1)_Y}
\newc{\smgg}{SU(3)_c\times SU(2)_L\times U(1)_Y}
\newc{\mtwo}{M_2}
\newc{\mone}{M_1}
\newc{\tanb}{\tan\beta}
\newc{\mw}{m_W}   \newc{\mz}{m_Z}
\newc{\mchi}{m_{\chi}}
\newc{\hinot}{\widetilde H^0_2}
\newc{\hinob}{\widetilde H^0_1}
\newc{\bino}{\widetilde B^0}
\newc{\wino}{\widetilde W^0_3}
\newc{\hinos}{\widetilde H_S}       \newc{\hinoa}{\widetilde H_A}
\newc{\mcharone}{m_{\charone}}	\newc{\charone}{\chi_1^\pm}
\newc{\gluino}{\widetilde g}
\newc{\mgluino}{m_{\gluino}}
\newc{\photino}{\widetilde\gamma}
\newc{\flr}{ f_{L,R} }
\newc{\sfermion}{\widetilde f}
\newc{\msf}{m_{\sfermion}}
\newc{\snu}{\widetilde\nu}
\newc{\sel}{\widetilde e}
\newc{\msl}{m_{\widetilde l}}
\newc{\msq}{m_{\widetilde q}}
\newc{\msel}{m_{\sel}}
\newc{\mhalf}{m_{1/2}}
\newc{\mnot}{m_0}		\newc{\mzero}{\mnot}
\newc{\azero}{A_0}	\newc{\bzero}{B_0}
\newc{\muzero}{\mu_0}

\newc{\mtop}{m_t}
\newc{\mbot}{m_b}
\newc{\mtau}{m_{\tau}}
\newc{\htop}{h_t}
\newc{\hbot}{h_b}
\newc{\htau}{h_{\tau}}

\newc{\sinsqtheta}{\sin^2\theta_w}
\newc{\alphas}{\alpha_{\rm s}}
\newc{\alphaem}{\alpha_{\rm em}}

\newc{\hone}{H_1}
\newc{\htwo}{H_2}
\newc{\vev}{{\it v.e.v.}}
\newc{\vone}{v_b}   \newc{\vtwo}{v_t}
\newc{\hl}{h}   \newc{\hh}{H}   \newc{\ha}{A}
\newc{\mhl}{m_\hl}   \newc{\mhh}{m_\hh}   \newc{\ma}{m_A}
    \newc{\ch}{C}   \newc{\chpm}{C^{\pm}}   \newc{\chmp}{C^{\mp}}
\newc{\mch}{m_\ch}   \newc{\mchpm}{m_\chpm}   \newc{\mchmp}{m_\chmp}
\def\NPB#1#2#3{Nucl. Phys. B {\bf#1} (19#2) #3}
\def\PLB#1#2#3{Phys. Lett. B {\bf#1} (19#2) #3}
\def\PLBold#1#2#3{Phys. Lett. {\bf#1B} (19#2) #3}
\def\PRD#1#2#3{Phys. Rev. D {\bf#1} (19#2) #3}
\def\PRL#1#2#3{Phys. Rev. Lett. {\bf#1} (19#2) #3}
\def\PRT#1#2#3{Phys. Rep. {\bf#1} C (19#2) #3}
\def\ARAA#1#2#3{Ann. Rev. Astron. Astrophys. {\bf#1} (19#2) #3}
\def\ARNP#1#2#3{Ann. Rev. Nucl. Part. Sci. {\bf#1} (19#2) #3}
\def\MODA#1#2#3{Mod. Phys. Lett. A {\bf#1} (19#2) #3}
\def\IJMPA#1#2#3{Int. J. Mod. Phys. A {\bf#1} (19#2) #3}
\def\APJ#1#2#3{Ap. J. {\bf#1} (19#2) #3}
\begin{flushright}
UM-TH-93-22\\
hep-ph/9309209\\
August 1993\\
\end{flushright}
\vskip 0.6in
\begin{center}
{\large \bf SUSY-GUT HIGGS PROBE\\
AT FUTURE $e^+e^-$ COLLIDERS
\footnote{\tenrm\baselineskip=11pt Talk at the {\em Workshop on  
Physics
and Experiments with Linear $e^+e^-$ Colliders}, Waikaloa, Hawaii,  
April
26-30, 1993, to appear in the Proceedings.}
\\}
\vskip .4in
{\large LESZEK ROSZKOWSKI}
\vskip .1in
{\em leszek@leszek.physics.lsa.umich.edu\\
     Randall Physics Laboratory,\\
     University of Michigan,\\
     Ann Arbor, MI 48109-1129, USA}
\end{center}
\vskip .2in
\begin{abstract}
\noindent
Several independent constraints on the minimal supersymmetric model  
with
constraints from grand unifications and cosmology select the region  
of
low-energy supersymmetry below ${\cal O}$(1~TeV). The resulting mass  
of the
lightest Higgs boson can probably
be covered at LEP II only if its beam energy exceeds 200~GeV, and is  
within the
detectability range of the NLC with $\sqrt{s}=300$~GeV.
\end{abstract}

%
     \setlength{\baselineskip}{14pt}
\setcounter{footnote}{0}
\setcounter{page}{1}	
\setcounter{section}{0}
\setcounter{subsection}{0}

\vskip 1cm
Supersymmetry and grand unifications have recently gained renewed  
interest
stemming from LEP measurements of the gauge couplings.
The running couplings do not meet at one point in the Standard Model,  
while
they do so in minimal supersymmetry~\cite{amaldi}.
This remarkable feature, along with its several other virtues, makes
supersymmetry a particularly attractive extension of the Standard  
Model.
Several of those desirable features
are simultaneously satisfied~\cite{dick,kkrw} in minimal  
supersymmetry for
energy scales below roughly 1 TeV. The resulting region of the
parameter space puts stringent constraints on the mass spectra
of supersymmetric particles. In this report~\cite{kkrw} I briefly  
describe the
resulting implications for the lightest Higgs boson searches at  
future
$e^+e^-$ colliders, like LEP II and the NLC.

I consider the minimal supersymmetric model
coupled to minimal supergravity which provides
the desired form of SUSY soft breaking terms.
The main unification assumptions of the minimal SUSY model coupled to  
minimal
supergravity are as follows:

\begin{enumerate}
\item
{\em Common gauge coupling $\gx$:} The three running gauge couplings  
meet at a
single energy scale $\mx$: $g_1(\mx)=g_2(\mx)=g_s(\mx)=\gx$, where
$g_1=\sqrt{3/5}g^\prime$.
\item
{\em Common gaugino mass $\mhalf$:} The soft SUSY breaking gaugino  
mass terms
satisfy $M_1(\mx)=M_2(\mx)=M_3(\mx)=\mhalf$.
\item
{\em Common scalar mass $\mzero$:} The soft SUSY breaking scalar mass  
terms
contributing to the squark, slepton, and Higgs masses are equal
to $\mzero$ at $\mx$.
\item
{\em Common trilinear scalar coupling $\azero$:} The soft trilinear  
SUSY
breaking terms are all equal
to $\azero$ at $\mx$.
\end{enumerate}
The bilinear soft scalar coupling $\bzero$ is sometimes assumed to be
related to $\azero$ via $\bzero$=$\azero-1$ but in general remains
an independent parameter.

In addition one often assumes that at $\mx$ the masses of the  
down-type quarks
and leptons of each generation are equal since in many GUTs they  
appear in the
same multiplets.
We find, however, that if we assume $\hbot(\mx)=\htau(\mx)$
but ignore GUT-scale threshold corrections then the resulting values
of $\mbot$ are some $20\%$ ($\sim 1$~GeV) too high~\cite{kkrw}.
On the other hand, even rather small (few per cent) GUT-scale  
corrections to
the running of $\hbot$ result in typically larger
shifts in $\mbot(\mbot)$~\cite{kkrw}. In light of this and since we  
do not
adhere to any specific GUT model, at this point we do not include  
this
constraint in limiting the allowed parameter space.
On the other hand, we do not find
the GUT-scale effects on the running of the gauge couplings and  
various mass
parameters equally important and we ignore them.

Thus in this model there are at least six fundamental quantities:
(a) the common gaugino mass $\mhalf$, (b) the common
scalar mass $\mnot$, (c) the Higgs/higgsino mass parameter $\mu$,
(d) the common scale $\azero\mzero$ of all the trilinear soft  
SUSY-breaking
terms,
(e) the bilinear soft mass parameter $\bzero\mzero$,
and (f) the top Yukawa coupling $\htop$.
Two minimization
conditions of the Higgs potential allow for expressing $\mu$ and  
$\bzero$ in
terms of $\tanb=\vtwo/\vone$ and $\mz$, and thus effectively
reducing the minimum number of independent parameters to five.
We take them to be: $\mzero$, $\mhalf$, $\azero$, $\tanb$, and  
$\mtop=
\htop\vtwo$. (There is an additional two-fold ambiguity in choosing  
the
sign of $\mu$.)
Since $\tanb$ can be large we also include $\hbot$ and $\htau$. The  
parameters
listed above determine the whole spectrum of masses and
couplings in the model.

The considered framework is remarkably predictive. As an input from
experiment we use only a few well measured quantities: $\mz$,  
$\alpha_{\rm
em}$, $\sinsqtheta(\mz)$, and, in considering $\mbot$, the
mass of the tau lepton. Since evaluation is usually done in the
$\msbar$-scheme, all these quantities must be `translated' into the
$\msbar$-scheme at specified energy scales.

By using the previously stated assumptions and the RG Equations  
(RGEs)
one is then able to calculate $\alphas(\mz)$, $\mbot$, and the masses  
of all
the squarks, sleptons, Higgs bosons, gluinos, charginos, and the  
neutralinos in
terms of a few basic parameters. One can then identify the lightest
supersymmetric
particle (LSP) and calculate its relic abundance. In addition we  
impose
dynamical gauge symmetry breaking yielding the correct value of  
$\mz$.
Finally, for each point in the parameter space one can estimate the  
amount of
fine-tuning needed to find phenomenologically interesting
solutions.

We simultaneously evolve all the relevant parameters between the GUT  
scale and
the electroweak scale
and properly take into account the effect of multiple mass thresholds  
on the
running of the gauge couplings. The details of the procedure can be  
found in
Ref.~\cite{kkrw}. I stress that we don't fix
a single SUSY breaking scale at some arbitrary value but rather
perform a global dynamical derivation of the low-energy mass spectra.
We also avoid making {\em ab initio} such potentially questionable
constraints on the parameter space as fixing the value of poorly  
measured
$\alphas$, requiring no fine-tuning, or assuming that
$m_{\widetilde q}\lsim$ 1 TeV.
Instead, we display the effect of various  constraints on the
{\em output} of the analysis. This way we have a much better control  
of
the meaning of our conclusions.
%
	\vspace{-3cm}
  \begin{center}
  \epsfxsize4in
  \hspace*{0in}
  \epsffile{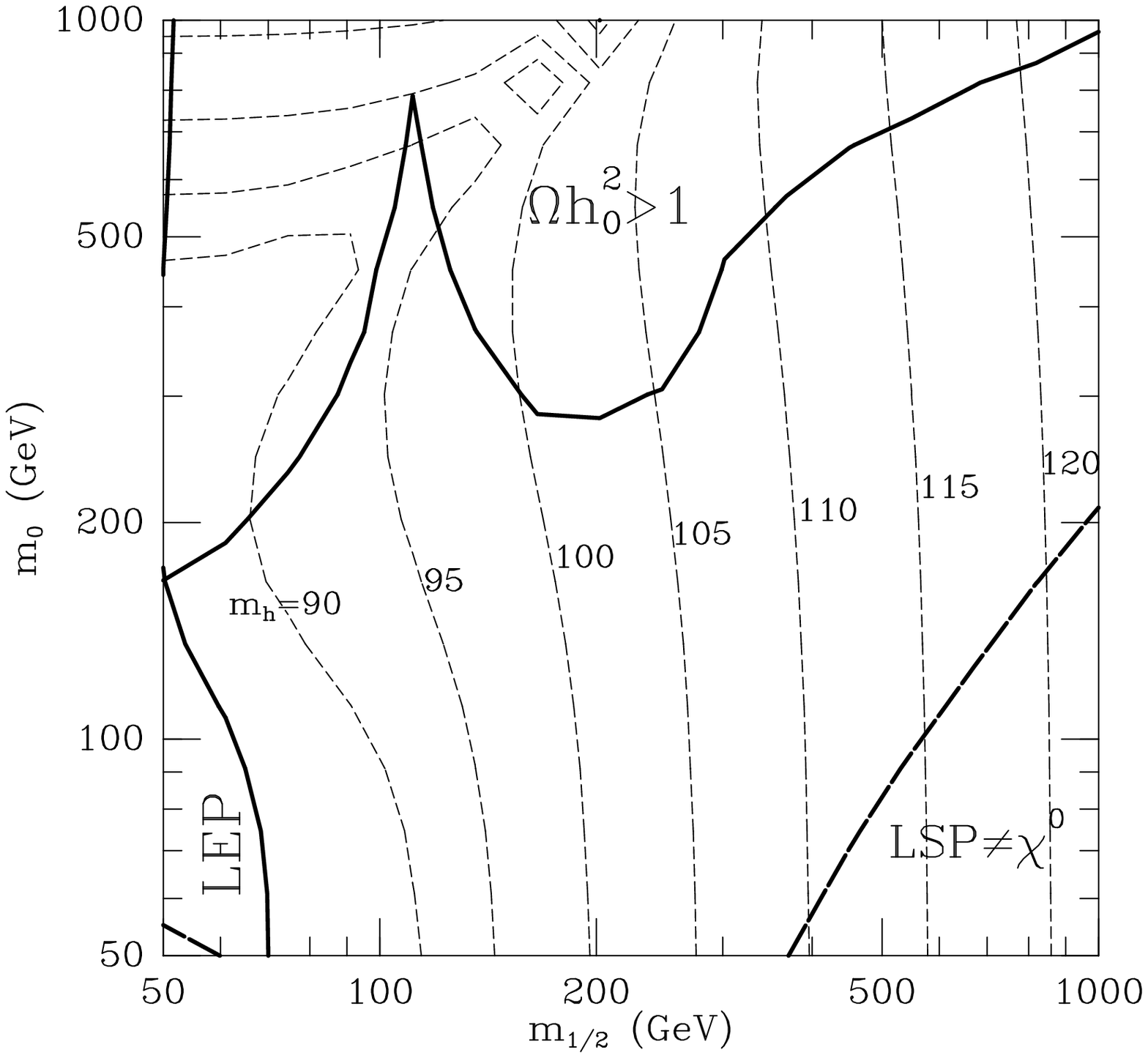}

\parbox{5.5in}{
\small Figure 1. In the plane
($\mhalf,\mnot$) for $\mtop=160$~GeV, $\tanb=5$, $\azero=3$, and  
$\mu>0$ I
present the mass contours of the Higgs boson $h$ (short-dashed line,  
in GeV).
Regions delineated by thick solid lines are excluded by: (a)
high-energy experiments (marked LEP in the Figure) (In this case it
comes from $m_{\chi^\pm}>46$~GeV.); and (b) the conservative lower  
bound on the
age of the
Universe of 10~Gyrs (marked $\Omega h^2_0>1$). The region marked
${\rm LSP}\neq\chi^0$ is not excluded but strongly disfavored by
cosmology (see text).
\label{figureone}
    }
\end{center}

I will now focus on the selected region of the parameter space.
This is presented in Fig.~1 for $\mtop=160$~GeV, $\tanb=5$,  
$\azero=3$, and
$\mu>0$. First, experimental constraints (in this case the most  
limiting being
$m_{\chi^\pm}\gsim46$~GeV, marked LEP) are at present extremely mild.  
Second,
the very powerful
cosmological constraint $\abund<1$ comes from a very conservative  
lower bound
of about 10 billion years on the age of the Universe~\cite{kt}. (It  
does not
depend on the specific nature (nor even the existence) of dark  
matter.) The
third constraint is not so firm: in the region marked  
`LSP$\neq\chi^0$' the
lightest
neutralino is not the LSP. Instead, it is either a sneutrino leading
to a very small relic abundance, $\Omega_{\widetilde\nu}\sim10^{-4}$,
or charged/colored sparticle (either stau or stop).
Both possibilities
are strongly disfavored on cosmological grounds but not firmly
excluded~\cite{kt}.
Finally, we somewhat arbitrarily limit $\mhalf<1$~TeV (which, via
$\Omega h^2_0$, puts a similar bound on $\mzero$). This corresponds  
to
$\mgluino\lsim2$~TeV (and similarly for the squarks and heavy Higgs  
bosons).
In the selected region $0.121<\alpha_s(\mz)<0.131$ (decreasing
slowly for $\mhalf,\mzero$ above 1~TeV), in an excellent agreement  
with
the experimental data~\cite{altarelli-eps}. (As I mentioned before,  
$\mbot$
comes out too large but the GUT-scale corrections are expected to be
significant~\cite{kkrw}.)
For other choices of parameters~\cite{kkrw} the relative importance  
of these
constraints varies, with the requirement of correct electroweak gauge  
symmetry
breaking often playing also a significant r{\^ole}, but typically  
they all
allow a broad region $\mzero\sim\mhalf$.

In Fig.~1 I also plot contours of the 1-loop-corrected Higgs boson  
mass
$m_h$ resulting from the analysis~\cite{kkrw}. In the selected region  
we find
$m_h\lsim120$~GeV and $h$ is SM Higgs-like. It is clear that the  
discovery
potential of LEP~II will crucially depend on its beam  
energy~\cite{kkrw}. If
one can reach
$\sqrt{s}$ above 200~GeV then a significant range of $m_h$ can be  
explored. In
contrast, chances of the Higgs discovery are less than slim for  
$\sqrt{s}$ in
the currently approved range of about 176~GeV, for which $m_h$ can be  
searched
up to about 80~GeV~\cite{lepstudies}. It is also clear that the NLC  
with even a
modest choice of $\sqrt{s}=300$~GeV will cover
the whole Higgs mass range in the expected region of $\mhalf$ and
$\mzero$~\cite{kkrw}. These conclusions apply also to other choices  
of
parameters. More detailed studies are needed to determine more  
precisely what
regions of the plane $\mhalf, \mzero$ will be covered
by Higgs searches as a function of $\sqrt{s}$ and luminosity.

Finally, I briefly mention about prospect for other SUSY particles
searches in this model~\cite{kkrw}. A priori they could be discovered  
even at
LEP~II and the Tevatron. However, requiring the LSP to be the  
dominant
component of (dark) matter in the Universe puts {\em lower} limits on
their masses making their discovery unlikely before the next  
generation
of supercolliders (SSC and LHC) and $e^+e^-$ linear colliders (NLC  
with
$\sqrt{s}=500 - 1000$~GeV)~\cite{dick,kkrw}. If so, then finding the  
lightest
Higgs may
be our window of opportunity for confirming supersymmetry before the
next millennium, in fact in just a few years.

\vskip 0.7cm
\section*{Acknowledgments}

\noindent
The presented work has been done in collaboration with G.~Kane,  
C.~Kolda, and
J.~Wells (Ref.~\cite{kkrw}).
\bigskip

\end{document}